\documentclass[twoside,9pt]{article}  % Remove 'twocolumn' for single column format
\usepackage{extsizes}
\usepackage[super,sort&compress,comma]{natbib}
\usepackage[version=3]{mhchem}
\usepackage[left=3.0cm, right=3.0cm, top=1.785cm, bottom=2.0cm]{geometry}
\usepackage{balance}
\usepackage{mathptmx}
\usepackage{sectsty}
\usepackage{graphicx}
\usepackage{lastpage}
\usepackage[format=plain,justification=justified,singlelinecheck=false,font={stretch=1.125,small,sf},labelfont=bf,labelsep=space]{caption}
\usepackage{float}
\usepackage{fancyhdr}
\usepackage{fnpos}
\usepackage[english]{babel}
\addto{\captionsenglish}{%
  \renewcommand{\refname}{Notes and references}
}
\usepackage{array}
\usepackage{droidsans}
\usepackage{charter}
\usepackage[T1]{fontenc}
\usepackage[usenames,dvipsnames]{xcolor}
\usepackage{setspace}
\usepackage[compact]{titlesec}
\usepackage{hyperref}

\usepackage{chemformula} % Formula subscripts using \ch{}
\usepackage{amsmath,amssymb,amsfonts}
\usepackage{bbding}
\usepackage[notextcomp]{stix}

\usepackage{epstopdf} % This line makes .eps figures into .pdf - please comment out if not required.

\definecolor{cream}{RGB}{222,217,201}

\begin{document}

\pagestyle{fancy}
\thispagestyle{plain}
\fancypagestyle{plain}{
%%% HEADER %%%
\renewcommand{\headrulewidth}{0pt}
}
%%% END OF HEADER %%%

%%% PAGE SETUP - Please do not change any commands within this section %%%
\makeFNbottom
\makeatletter
\renewcommand\LARGE{\@setfontsize\LARGE{15pt}{17}}
\renewcommand\Large{\@setfontsize\Large{12pt}{14}}
\renewcommand\large{\@setfontsize\large{10pt}{12}}
\renewcommand\footnotesize{\@setfontsize\footnotesize{7pt}{10}}
\makeatother

\renewcommand{\thefootnote}{\fnsymbol{footnote}}
\renewcommand\footnoterule{\vspace*{1pt}%
\color{cream}\hrule width 3.5in height 0.4pt \color{black}\vspace*{5pt}}
\setcounter{secnumdepth}{5}

\makeatletter
\renewcommand\@biblabel[1]{#1}
\renewcommand\@makefntext[1]%
{\noindent\makebox[0pt][r]{\@thefnmark\,}#1}
\makeatother
\renewcommand{\figurename}{\small{Fig.}~}
\sectionfont{\sffamily\Large}
\subsectionfont{\normalsize}
\subsubsectionfont{\bf}
\setstretch{1.125} % In particular, please do not alter this line.
\setlength{\skip\footins}{0.8cm}
\setlength{\footnotesep}{0.25cm}
\setlength{\jot}{10pt}
\titlespacing*{\section}{0pt}{4pt}{4pt}
\titlespacing*{\subsection}{0pt}{15pt}{1pt}
%%% END OF PAGE SETUP %%%

\newenvironment{psmallmatrix}
  {\left(\begin{smallmatrix}}
  {\end{smallmatrix}\right)}

%%% FOOTER %%%
\fancyfoot{}
\fancyfoot[LO,RE]{\vspace{-7.1pt}\includegraphics[height=9pt]{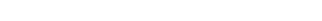}}
\fancyfoot[CO]{\vspace{-7.1pt}\hspace{13.2cm}\includegraphics{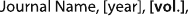}}
\fancyfoot[CE]{\vspace{-7.2pt}\hspace{-14.2cm}\includegraphics{head_foot/RF}}
\fancyfoot[RO]{\footnotesize{\sffamily{1--\pageref{LastPage} ~\textbar  \hspace{2pt}\thepage}}}
\fancyfoot[LE]{\footnotesize{\sffamily{\thepage~\textbar\hspace{3.45cm} 1--\pageref{LastPage}}}}
\fancyhead{}
\renewcommand{\headrulewidth}{0pt}
\renewcommand{\footrulewidth}{0pt}
\setlength{\arrayrulewidth}{1pt}
\setlength{\columnsep}{6.5mm}
\setlength\bibsep{1pt}
%%% END OF FOOTER %%%

%%% FIGURE SETUP - Please do not change any commands within this section %%%
\makeatletter
\newlength{\figrulesep}
\setlength{\figrulesep}{0.5\textfloatsep}

\newcommand{\topfigrule}{\vspace*{-1pt}%
\noindent{\color{cream}\rule[-\figrulesep]{\columnwidth}{1.5pt}} }

\newcommand{\botfigrule}{\vspace*{-2pt}%
\noindent{\color{cream}\rule[\figrulesep]{\columnwidth}{1.5pt}} }

\newcommand{\dblfigrule}{\vspace*{-1pt}%
\noindent{\color{cream}\rule[-\figrulesep]{\textwidth}{1.5pt}} }

\makeatother
%%% END OF FIGURE SETUP %%%

%%% TITLE, AUTHORS AND ABSTRACT %%%
\sffamily
\begin{center}
  \LARGE{\textbf{Tunable Assembly of Confined Janus Microswimmers in Sub-kHz AC Electric Fields under Gravity}} \\ % Article title
  \vspace{0.3cm} 
  \large{Carolina van Baalen,$^{a}$ Laura Alvarez,$^{b}$ Robert Style,$^{a}$ Lucio Isa$^{a \ast}$} \\ % Author names
  \vspace{0.3cm}
  \begin{minipage}{0.8\textwidth} % Adjust the width as needed
    \normalsize{
    Active systems comprising micron-sized self-propelling units, 
    also termed microswimmers, are promising candidates for the 
    bottom-up assembly of small structures and reconfigurable materials. 
    Here we leverage field-driven colloidal assembly to induce 
    structural transformations in dense layers of microswimmers 
    driven by an alternating current (AC) electric field and confined 
    in a microfabricated trap under the influence of gravity. 

    By varying the electric field frequency, we realize significant 
    structural transformations, from a gas-like state at high frequencies 
    to dynamically rearranging dense crystalline clusters at lower 
    frequencies, characterized by vorticity in their dynamics. 

    We demonstrate the ability to switch between these states on-demand, 
    showing that the clustering mechanism differs from motility-induced 
    phase separation. Our results offer a valuable framework for designing 
    high-density active matter systems with controllable structural properties, 
    envisioned to advance the development of artificial materials with 
    self-healing and reconfiguration capabilities.
    }
  \end{minipage}
\end{center}

\vspace{1cm} % Add some space before footnotes

%%% FOOTNOTES %%%

\footnotetext{\textit{$^{a}$~Laboratory for Soft Materials and Interfaces, Department of Materials, ETH Zürich, Vladimir-Prelog-Weg 5, 8093 Zürich, Switzerland. Email: lucio.isa@mat.ethz.ch}}
\footnotetext{\textit{$^{b}$~Univ. Bordeaux, CNRS, CRPP, UMR 5031, F-33600 Pessac, France, Email: laura.alvarez-frances@u-bordeaux.fr}}

%%%MAIN TEXT%%%%
\section{Introduction}
Unlike steady-state equilibrium or driven matter, active systems comprise autonomous units that constantly self-propel by locally converting energy from the imbalance of a specific thermodynamic quantity, such as temperature, chemical or ionic concentration, yielding a range of distinctive features. The collective translation of such self-propelling units can lead to the formation of coherent patterns and self-organization — an essential characteristic for the proper functioning of many biological systems across various length scales, ranging from the antipredatory function of large bird flocks \cite{Zoratto2009}, down to the enhanced colonization abilities of swarming motile bacteria \cite{Lai2009}, and the dynamic actin-myosin cytoskeleton critical to numerous aspects of cell physiology \cite{Olson2022}. Particularly interesting from a physics as well as material science perspective is the study of microscopic active matter, which has uncovered a variety of new behaviors unattainable in equilibrium systems. 

Microscopic active systems can display non-trivial density profiles and critical fluctuations, even at densities much lower than the bulk jamming density found in passive systems. In such systems, the self-propelling units, often referred to as microswimmers or active particles, may undergo dynamic clustering akin to a liquid-gas phase separation \cite{Bialke2013, Thompson2011, Levis2014, Redner2013, Fily2012, Speck2014, Cates2015} or crystallization \cite{Tan2022, Petroff2015, Chen2015, Petroff2018, Palacci2013, Bililign2022, Buttinoni2013} at relatively low particle densities and in the absence of any temperature changes or attractive inter-particle interactions, making microswimmers promising candidates for the bottom-up assembly of small structures and reconfigurable materials that can change their properties on demand. 

Numerical studies have shown that microswimmer self-assembly can arise through self-trapping driven by the interplay of out-of-equilibrium collisions and finite reorientation times. This mechanism is fundamentally governed by microswimmer density, self-propulsion velocity, and rotational diffusion \cite{Tailleur2008, Fily2012, Redner2013}. Although attractive inter-particle interactions are not required, they can facilitate the onset of collective motion and expand the range of dynamic structures achievable \cite{Redner2013, Martin2021, Sese2022}. Indeed, different experiments have demonstrated the existence of various collective states of synthetic microswimmers, including polar liquid states \cite{Bricard2013, Driscoll2017}, turbulence \cite{Karani2019}, swarms \cite{Yan2016}, and living crystals \cite{Palacci2013, Buttinoni2013}. Nonetheless, experimentally achieving a wide range of on demand controlled structures remains a challenging task, in particular, due to the need for a single experimental system with controlled microswimmer densities, as well as in-situ control over the microswimmers' self-propulsion speed and interaction potential. 

In this manuscript, we utilize gravity and electric field-driven colloidal assembly to achieve controlled structural organization and self-propulsion dynamics in dense layers of colloidal microswimmers. By varying the frequency of the applied electric field, we induce structural transitions spanning gas-like states to self-assembled motile crystallites, accompanied by emergent vorticity at lower field frequencies. Furthermore, we demonstrate the ability to switch between these states on demand by modulating the field frequency. This approach enables the creation of a high-density active matter system with strong, tunable interactions, distinct from prior studies emphasizing assembly driven primarily by motion persistence \cite{Buttinoni2013, Palacci2013}.

\section{Results} 
\subsection{Confined active colloidal monolayers under gravity}
\begin{figure*}[!ht]
    \centering
    \includegraphics[scale=.48]{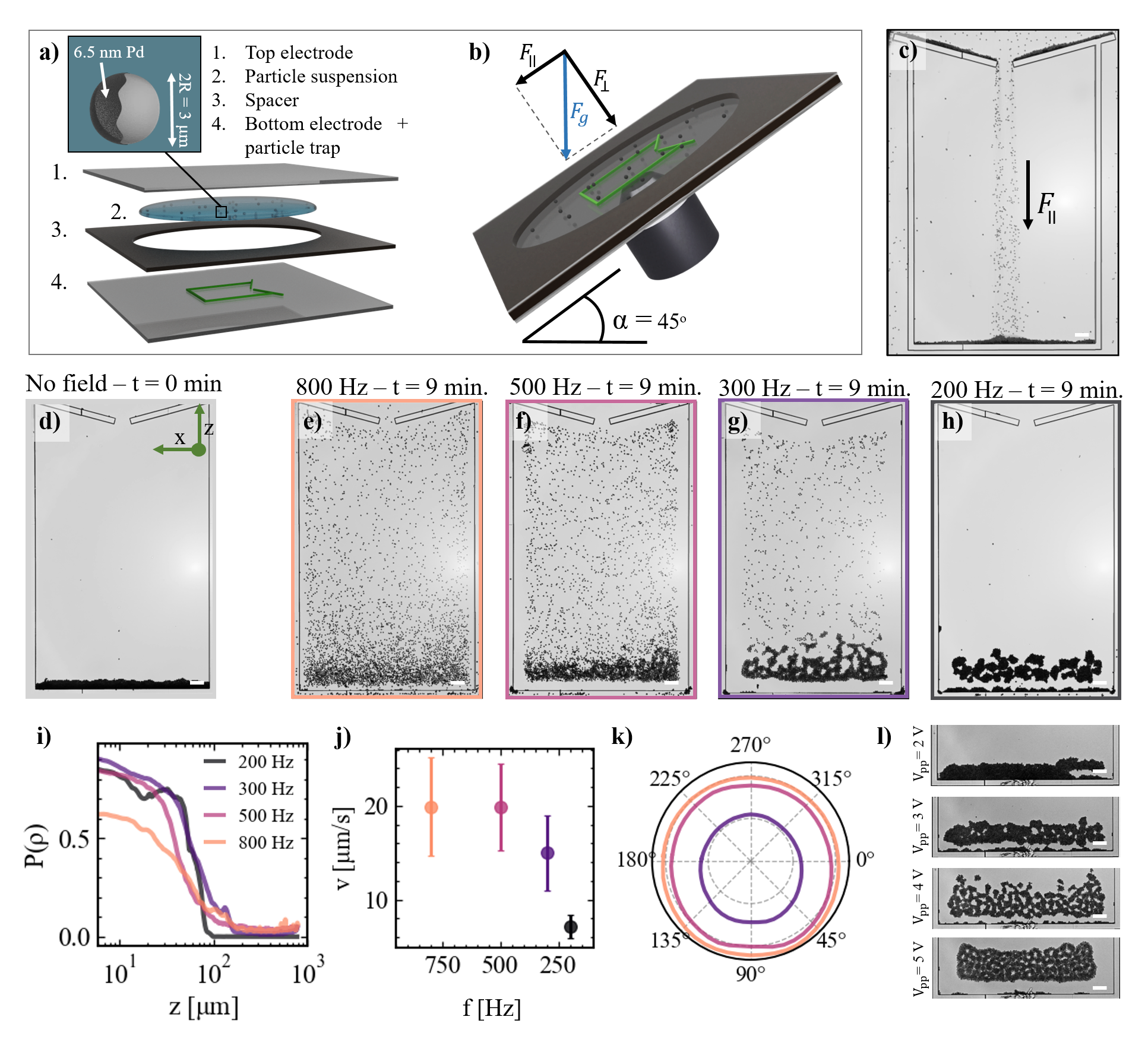}
    \caption{\textbf{AC field-propelled Janus particles swimming against gravity at different frequencies. } a) Illustration (not to scale) showing the composition of the experimental cell. Details of the Janus microswimmers are shown in the inset. b) Experimental setup. The entire setup is tilted at an angle of $\alpha$=45$^\circ$, resulting in a gravitational force ($F_g$) perpendicular (F$_\perp$) and parallel (F$_\parallel$) to the laboratory frame. c) Optical microscopy image of the Janus particles sedimenting into the microfabricated trap under the action of F$_\parallel$ in the absence of the electric field. d) Optical microscopy image of a filled trap after $\sim$3 hours of sedimentation in the absence of the electric field. e-h) Snapshots of the system after 9 min. of equilibration at different AC electric field frequencies at a peak-to-peak voltage of $V_{pp}$ = 4 V, and corresponding i) mean density profiles as a function of height $z$ in the trap. j) Mean microswimmer self-propulsion velocities extracted from the mean squared displacements (MSDs) on a horizontal substrate ($\alpha$=0$^\circ$, error bars represent the standard deviation from at least 20 particles), and k) microswimmer velocities as a function of orientation ($\alpha$=45$^\circ$), calculated from displacements over a 33 msec. time interval in the headspace of the particle trap. l) Effect of changing the applied peak-to-peak voltage from $V_{pp}$=2 to 5 V on the structural properties of the active monolayer formed at a fixed frequency of 200 Hz. The color code in i-k) corresponds to frame colors in e-h). Scale bars: 50 $\mu$m. 
    }
    \label{fig:Setup_GlobalFeatures}
\end{figure*}

We examine the behavior of metallo-dielectric Janus spheres confined in an inclined microfabricated chamber under an applied electric field, focusing on how frequency modulation affects their structural organization and dynamics. Our samples comprise a suspension of Pd-capped SiO$_2$ Janus spheres (diameter 2R=3 $\mu$m) residing between two parallel electrodes held at a 30 $\mu$m distance using a spacer (Fig. \ref{fig:Setup_GlobalFeatures}a). The bottom electrode integrates a microfabricated rectangular 1.0x0.6 mm$^2$ (LxW) chamber with a funneling inlet at the top (details on the sample preparation can be found in \ref{C4:SamplePrep}). Once assembled, we place the experimental cell at an angle of $\alpha = 45^\circ$, yielding a gravitational force perpendicular (F$_\perp$) and parallel (F$_\parallel$) to the laboratory frame (Fig. \ref{fig:Setup_GlobalFeatures}b). Under the action of gravity, the particles slide on the bottom electrode and down into the opening of the microfabricated chamber (Fig. \ref{fig:Setup_GlobalFeatures}c). After approximately 3 hours, denoted as t=0, a dense sediment has formed at the bottom of the chamber (Fig. \ref{fig:Setup_GlobalFeatures}c). Over the course of the entire experiment, we keep the experimental cell at $\alpha = 45^\circ$, which not only allows us to collect the microswimmers before turning the activity on, but also enables studying high-density active monolayers under slight compression. We connect the top and bottom electrodes to an electric field generator, producing an AC field with a chosen amplitude and frequency in the direction perpendicular to the electrodes. The rapidly alternating field causes a partial buildup of the ionic double layer near the electrodes and simultaneously polarizes the metallo-dielectric Janus spheres, yielding a gradient in the electric field where the particle meets the electrode. The latter causes the formation of electrohydrodynamic (EHD) flows\cite{Squires2006, Ristenpart2007}, which are of different magnitude and direction on the Pd side of the Janus particle, i.e. due to better polarizability of Pd compared to SiO$_2$, resulting in the propulsion of the Janus spheres with their SiO$_2$ face in front. As we will describe later, the frequency of the applied AC field also affects interparticle interactions mediated by EHD flows. 

We allow the active system to evolve for 9 min. under a fixed applied peak-to-peak voltage of $V_{pp}$ = 4 V and varying frequency from $f$=800 Hz to $f$=200 Hz (Supplementary Videos S1-S4). Note that the boundaries of the chamber are strongly repulsive upon turning on the AC field, therefore particles do not enter a region of approximately 30 $\mu$m from the walls of the chamber, hence they cannot escape through the top hole over the course of the experiment, and the chamber thus effectively acts as a trap. As shown qualitatively in Fig. \ref{fig:Setup_GlobalFeatures}e-h, applying different electric field frequencies results in markedly distinct structural features of the active colloidal monolayers. At a frequency of $f$=800 Hz, the microswimmers explore the entire space of the trap, with an average probability density $P(\rho)$ as a function of height $z$ that decays gradually from the bottom of the trap to the top (Fig. \ref{fig:Setup_GlobalFeatures}i). At $f$=500 Hz, the structure of the microswimmers' monolayer appears similar, but with a higher particle density at the bottom of the trap. Decreasing the frequency further to $f$=300 Hz results in increasing particle densities at the bottom, and finally at $f$=200 Hz only a dense sediment is left at the bottom of the trap, while no free swimming particles remain in the head space, as quantitatively apparent from the sharp drop in $P(\rho)$ versus $z$. To understand if the changing structure of the monolayers can be a result of changes in the self-propulsion velocities of the microswimmers, we measure their free swimming velocities in the absence of gravity, i.e. above a horizontal substrate (i.e. $\alpha = 0^\circ$), as shown in Figure \ref{fig:Setup_GlobalFeatures}j. Upon decreasing the frequency of the electric field from $f$=800 Hz to $f$=500 Hz, the mean self-propulsion velocities of the microswimmers remain approximately constant at around $v\approx$ 20 $\mu$m/s. However, further reducing the frequency from $f$=500 Hz to $f$=200 Hz, causes a marked decline in the mean self-propulsion velocity to $v\approx $7 $\mu$m/s. The sedimentation force acting on a single microswimmer, as approximated from the buoyant weight of a 3 $\mu$m SiO$_2$ sphere, is $F_g\approx$0.22 pN in our experiment. In contrast, the observed self-propulsion velocities at frequencies $f$=800, 500 and 300 Hz correspond to active forces, obtained from Stokes' drag, of $F_s \approx$ 0.51, 0.49, and 0.37 pN, respectively. 
Thus, at these frequencies, the average self-propulsion force is sufficient to enable microswimmers to overcome gravitational forces and swim upwards. A minor effect of gravity becomes discernible at these frequencies, as evidenced in the polar plot in Figure \ref{fig:Setup_GlobalFeatures}k. Here, we show the mean instantaneous velocities measured in the head space of the trap (i.e., at $\alpha = 45^\circ$) as a function of the swimming direction for $f$=800, 500 and 300 Hz, revealing self-propulsion velocities approximately $\sim$3 $\mu$m/s higher in the downward direction compared to the upward direction. Conversely, at $f$=200 Hz, the average self-propulsion force of the microswimmers falls below the threshold needed to swim upwards ($F_s \approx$ 0.18 pN), which explains the absence of microswimmers in the headspace of the trap at this frequency. The pronounced changes in the suspension microstructure with decreasing field frequency, however, cannot be solely attributed to variations in microswimmer persistence, as we will discuss in Section 2.2-3.

\begin{figure*}[!ht]
    \centering
    \includegraphics[scale=.6]{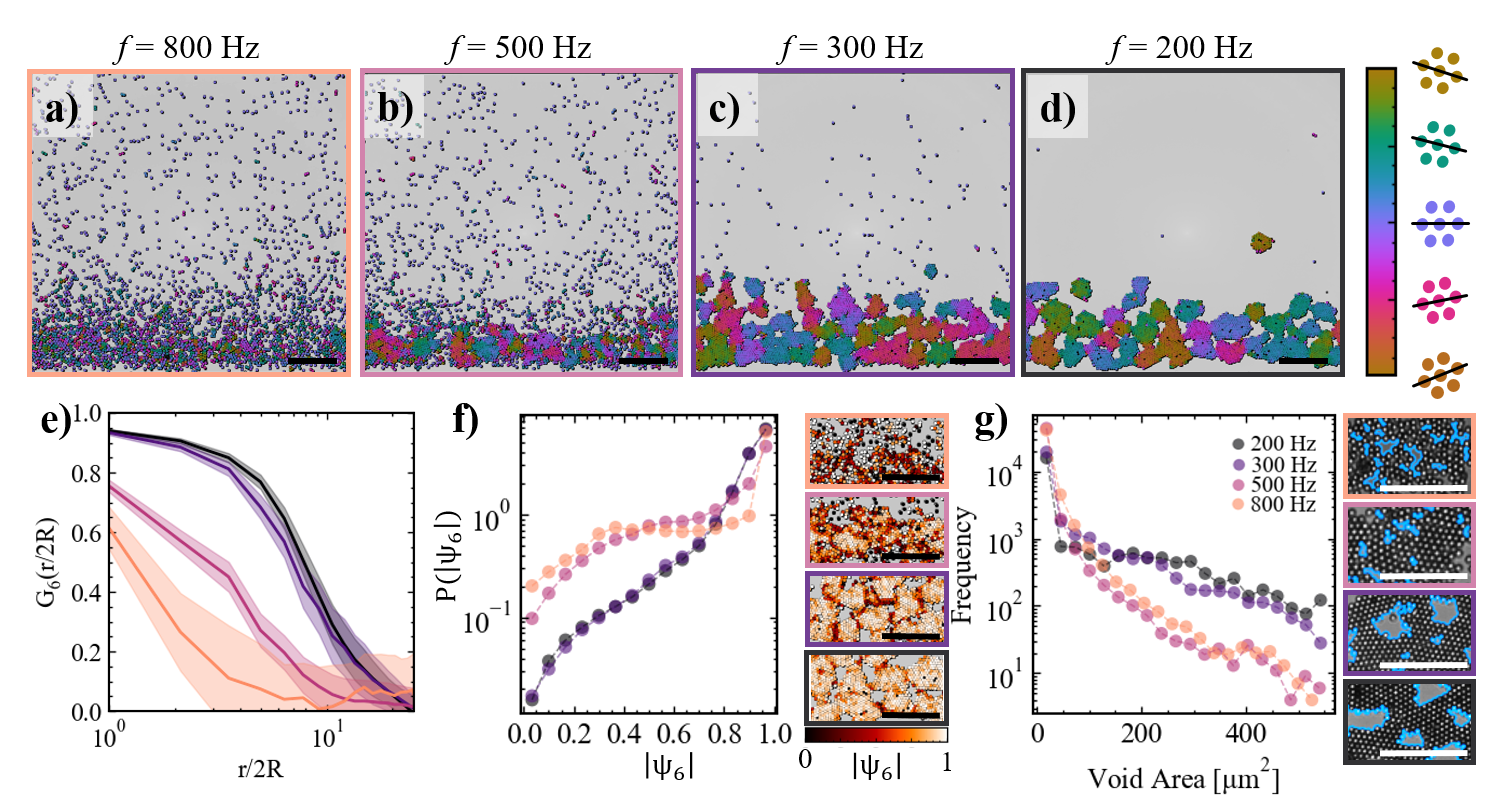}
    \caption{\textbf{Static structural information of microswimmer monolayers at different frequencies. } a-d) Snapshots of the system after 9 min. of equilibration at different frequencies at $V_{pp}$ = 4 V. Particles are color-coded according to their local orientation.  The frame color-coding corresponds to Fig. 1 e-h. e) Correlation function of the local orientation versus distance normalized by the particle size for the different $f$. f) Distribution of the absolute value of the hexagonal order parameter $\psi_6$ (left) and typical snapshots, with particles color-coded according to the absolute value of $\psi_6$. g) Distribution of void sizes (left) and typical snapshots of identified voids (right). Scale bars: 50 $\mu$m. 
    }
    \label{fig:StructuralFeatures}
\end{figure*}

We note that changing the applied field strength provides an additional handle to modify the structural properties of the active monolayers, as shown in Fig.  \ref{fig:Setup_GlobalFeatures}l. Increasing the applied peak-to-peak voltage from $V_{pp}$ = 2 to 5 V at a fixed frequency of $f$=200 Hz slightly decreases the overall density of the sediment and yields an increasing number of voids within it. We also observe an increasing mobility of the particles within the structures upon increasing the applied peak-to-peak voltage, which is likely responsible for the less compact structure observed at $V_{pp}$ = 5 V compared to $V_{pp}$ = 4 V. Nevertheless, in what remains, we will focus on the effect of modulating the frequency of the electric field, which allows for a greater range of distinct structural features.

\subsection{Structural features}
Having described the qualitative features of our system, we start by quantitatively investigating its structural properties. At the highest frequency, i.e. $f$=800 Hz, the microswimmers are dispersed throughout the entire trap, and the system shows features analogous to the active gas-like state observed for suspension of Pt-capped catalytically active Janus particles under a gravity field \cite{Palacci2010,Ginot2015, Ginot2018}. Notably, after 9 min. of equilibration at 800 Hz (\ref{fig:Setup_GlobalFeatures}i), the average density profile in the center of the particle trap decays exponentially with height, analogous to refs. \cite{Palacci2010,Ginot2015, Ginot2018} and theoretically discussed by \cite{Vachier2019, Enculescu2011, Szamel2014} (Supporting Information S1). Moreover, throughout the entire trap, the structure of the suspension is disordered, as visually apparent from zoomed-in snapshots of the active gas-like monolayer with particles color coded according to their local orientation $\psi_6$ (Fig. \ref{fig:StructuralFeatures}a). This is further quantitatively displayed in Figure\ref{fig:StructuralFeatures}e, which shows a spatial correlation of the local orientation $G_6(r/2R)$ that decays over length scales on the order of the particle diameter (orange data). Consequently, the sixfold symmetry in the system is low, characterized by the high probability of low values of the absolute value of the hexagonal order parameter |$\psi_6$| (Fig. \ref{fig:StructuralFeatures}f). Upon reducing the frequency of the AC electric field to $f$=500 Hz, clear differences appear in the structural properties of the system. As shown in Figure \ref{fig:StructuralFeatures}b, clusters emerge with orientations that remain correlated over larger length scales (Fig. \ref{fig:StructuralFeatures}e). Nonetheless, the overall order of the system remains low, as shown by the corresponding probability distribution of the absolute value of the hexagonal order parameter (Fig. \ref{fig:StructuralFeatures}f). Instead, at even lower frequencies, i.e. $f$=300 Hz and $f$=200 Hz, the probability distributions of the absolute value of the hexagonal order parameter show a shift towards higher values. Additionally, the color-coded snapshots in Figure \ref{fig:StructuralFeatures}c and d show manifestation of crystalline clusters, i.e. crystallites, comprising hexagonally assembled particles with strong orientational order that appear structurally similar to the "living crystals" observed in experiments focusing on the assembly of photocatalytic and thermophoretic microswimmers, which are formed through motion persistence \cite{Palacci2013, Buttinoni2013}. Moreover, aside from the few single free-swimming microswimmers at $f$=300 Hz, the structures formed at the bottom of the trap at $f$=300 Hz and $f$=200 Hz are essentially entirely composed of connected crystallites comprising hexagonally packed microswimmers. Consequently, $G_6$ only starts to decay significantly after $\sim$ 10 particle diameters, indicating the characteristic size of the crystalline domains at low $f$. Finally, associated with the emergence of the crystallites, voids correspondingly emerge in the structures at $f$=300 Hz and $f$=200 Hz, as highlighted in Figure \ref{fig:StructuralFeatures}g. Namely, upon decreasing $f$, the distributions of void sizes shift from mainly containing small voids with sizes <50 $\mu$m at $f$=800 Hz towards including void sizes >500 $\mu$m at $f$=300 Hz and $f$=200 Hz. 

\begin{figure*}[t!]
    \centering
    \includegraphics[scale=.6]{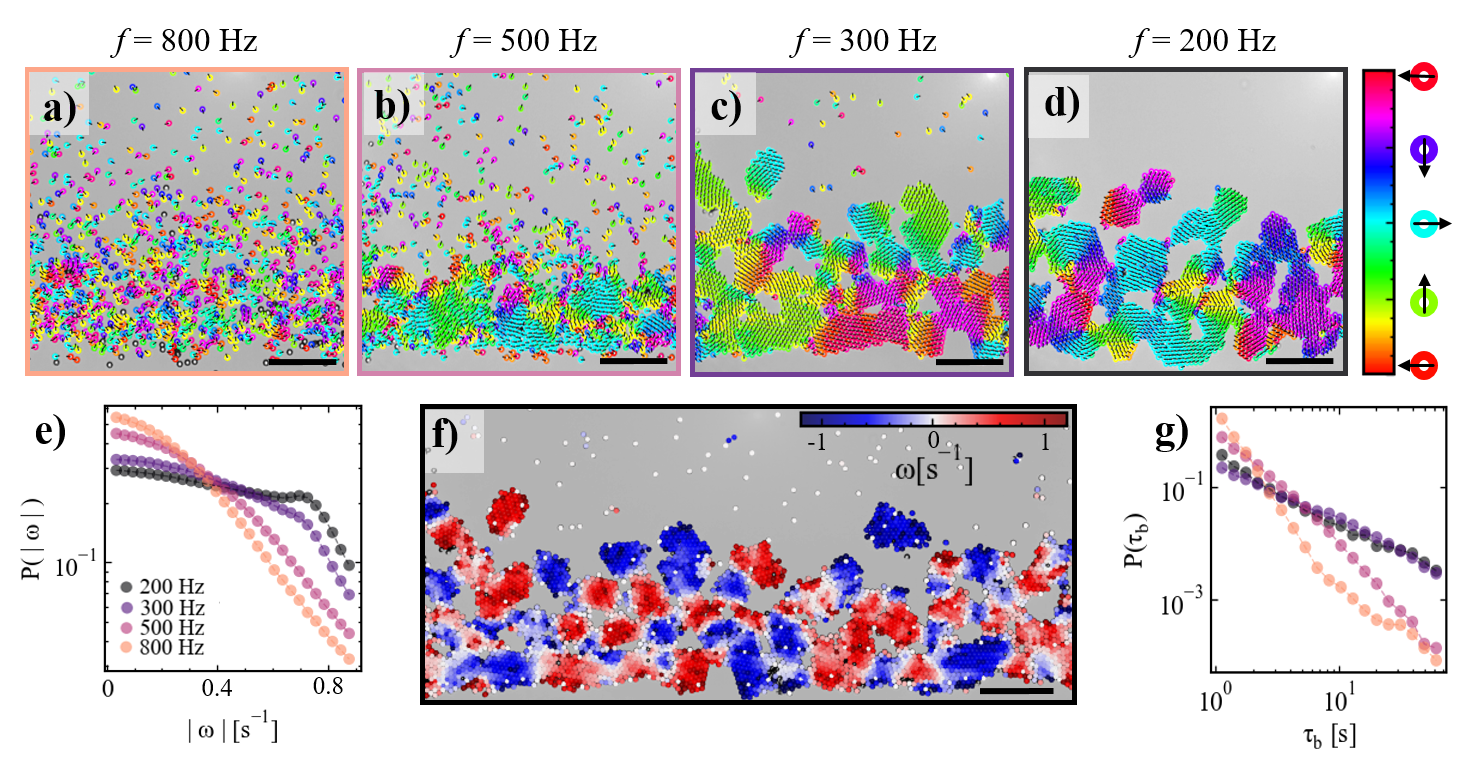}
    \caption{\textbf{Dynamics of microswimmer monolayers: the presence of vorticity. } a-d) Snapshots of the system after 9 min. of equilibration at different frequencies at $V_{pp}$ = 4 V, particles are color-coded according to their direction of motion extracted from the velocity vector. The arrows indicating the direction of motion in b-d) show the presence of vortices. Frame colorcoding corresponds to Fig. 1 a-d. e) Probability distributions of the absolute vorticity at different frequencies (See section \ref{C4:MM_DataAnalysis} for details on the calculation). f) Snapshot of a larger field of view corresponding to c), color-coded according to the local vorticity values. g) Probability distributions of the bond duration of particles at different frequencies. Note that for lower frequencies bond durations exceed the timescale of the experiment, hence the true distribution is not fully displayed. Scale bars: 50 $\mu$m. 
    }
    \label{fig:DynamicalFeatures}
\end{figure*}

\subsection{Dynamical features}
Having examined the static structural features of our system, we now turn our attention to their dynamic properties, exploring how microswimmer motion and clustering behavior evolve with decreasing electric field frequency. At a frequency of $f=800$ Hz, in the gas-like state, the direction of motion of individual microswimmers appears uncorrelated, as visually apparent in Figure \ref{fig:DynamicalFeatures}a, where particles are color-coded according to their direction of motion.

Decreasing the frequency to $f=500$ Hz leads to the emergence of collective motion within the system. Specifically, Figure \ref{fig:DynamicalFeatures}b illustrates vortex-like dynamics at this frequency, marking a significant shift from the uncorrelated motion observed at $f=800$ Hz. Similarly, at lower frequencies ($f=300$ Hz, Fig. \ref{fig:DynamicalFeatures}c, and $f=200$ Hz, Fig. \ref{fig:DynamicalFeatures}d), the crystallites in the system also exhibit vortex-like behavior, indicating the prevalence of rotational dynamics across these frequencies.

To quantify this behavior, we calculate the vorticity $\omega$ [s$^{-1}$] of the microswimmers at different frequencies and analyze the probability distribution of the absolute vorticity (Fig. \ref{fig:DynamicalFeatures}e). At $f=800$ Hz, the probability distribution exhibits a sharp peak at $|\omega|=0$, which gradually decays to zero at approximately $|\omega| \approx 0.9$ s$^{-1}$. Reducing the frequency to $f=500$ Hz slightly decreases the peak probability at $|\omega|=0$ and extends the tail of the distribution. At $f=300$ Hz, this extension becomes significantly more pronounced, reflecting an increase in rotational activity. Further lowering the frequency to $f=200$ Hz results in the emergence of a secondary peak around $|\omega| \approx 0.7$ s$^{-1}$, marking a substantial enhancement of rotational motion within the system.

Remarkably, the state of our system at frequencies $f \leq 300$ Hz shares similarities with previous experiments involving active chiral Quincke rollers \cite{Zhang2020, Zhang2022}, magnetically rotating colloidal rods \cite{Mecke2023}, and spinning starfish embryo crystals \cite{Tan2022}. However, unlike those systems, the active agents in our study lack intrinsic chirality or spinning behavior. The spontaneous vortical dynamics of hexagonally packed aggregates comprising randomly oriented Janus particles has been numerically predicted using a pusher-type squirmer model \cite{Shen2019}, where vorticity arises from microswimmer orientation and the geometry of the hexagonally packed aggregates. Figure \ref{fig:DynamicalFeatures}f shows a representative spatial map of vorticity for the low-frequency case, highlighting regions of correlated vorticity associated with crystallites rotating almost rigidly. Neighboring crystallites tend to rotate in opposite directions, as evidenced by the alternating blue and red regions in the figure. This cogwheel-like behavior likely results from hydrodynamic interactions, where a rotating crystallite's flow field drags its neighbor into phase-locked counter-rotation, as also reported in \cite{Shen2019}.

To further assess the rigidity of these structures, we analyze the distribution of bond durations, defined as the duration of geometrical contact between neighboring particles, across different frequencies (Fig. \ref{fig:DynamicalFeatures}g). At higher frequencies ($f=800$ Hz and $f=500$ Hz), bonds are predominantly short-lived, with the distribution falling to zero within the 1-minute measurement time frame. As the frequency decreases, the average bond duration increases markedly. At $f=300$ Hz and $f=200$ Hz, a significant fraction of bonds persist beyond the measurement time frame, highlighting the emergence of stable, coordinated assemblies that remain intact over extended periods.

\begin{figure*}[!ht]
    \centering
    \includegraphics[scale=.63]{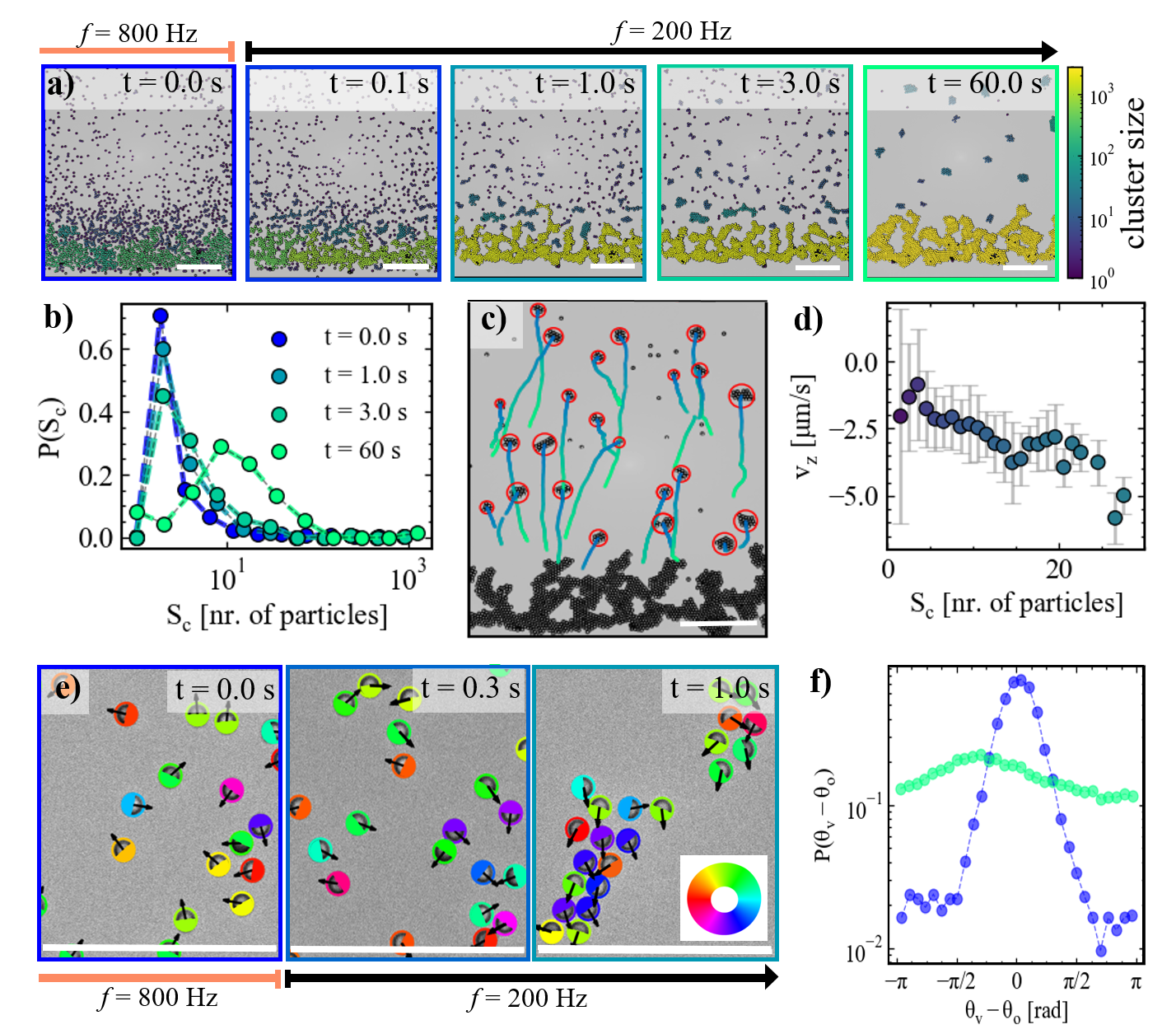}
    \caption{\textbf{Condensation of microswimmers 
    under gravity by switching the frequency of the AC electric field. } a) Snapshots of the system at different times after switching the frequency from 800 Hz to 200 Hz. Particles are color-coded according to the size of the cluster they belong to. b) Cluster size distributions at 0, 1, 3, and 60 seconds after switching the AC field frequency from 800 Hz to 200 Hz. c) Trajectories of clusters formed upon switching the frequency of the electric field (increasing time from blue to green). d) Sedimentation speed of clusters as a function of their size. Error bars represent the standard deviation over at least 20 measurements. f) High magnification snapshots of the system at different times after switching the frequency from 800 Hz to 200 Hz. Particles are color-coded according to their orientation, with the cap artificially colored. Black arrows indicate the direction of the velocity vector for each particle. g) Probability distributions of the difference between the cap orientation and the orientation of the velocity vector 5 seconds before (blue) and 5 seconds after (mint green) switching the AC electric field. Scale bars: 50 $\mu$m. 
    }
    \label{fig:Condensation}
\end{figure*}

\subsection{On-demand crystallisation}
A key feature of the system is its ability to undergo on-demand reconfiguration by adjusting the frequency of the AC electric field. In Figure \ref{fig:Condensation}, we demonstrate the "condensation" of an active colloidal monolayer by rapidly lowering the frequency from $f$=800 Hz to $f$=200 Hz. Figure \ref{fig:Condensation}a shows the immediate evolution of the monolayer after this frequency drop. Following the frequency reduction, small clusters form almost instantaneously and grow over time, as reflected by the cluster size distributions $P(S_c)$ shifting to larger values (Fig. \ref{fig:Condensation}b). Clusters initially formed in the upper region of the trap sediment downward, as shown by their trajectories in Figure \ref{fig:Condensation}c. The sedimentation speed increases with the growing size of the clusters (Fig. \ref{fig:Condensation}d), and they continue to fall until they merge with the fully connected structure at the bottom of the trap. After approximately 1.5 minutes, the resulting monolayer exhibits similar structural and dynamical features as those seen in systems equilibrated for 9 minutes at $f$=200 Hz.

To gain a deeper understanding of the particle-level mechanisms driving cluster formation, we analyze the orientation of the microswimmer caps (Supplementary Video S5). In Figure \ref{fig:Condensation}e, high-magnification images show the metallic caps of the Janus microswimmers color-coded by their orientations ($\theta_o$), with the direction of motion indicated by black arrows extracted from the velocity vector ($\theta_v$). At $f$=800 Hz (first panel, t=0.0 s), interparticle interactions are minimal, and the difference between $\theta_o$ and $\theta_v$ is small, as seen in the black arrows that are mostly perpendicular to the caps. This is corroborated by the narrow normal distribution of $\theta_v-\theta_o$ centered around zero (dark blue data in Figure \ref{fig:Condensation}f). After switching the frequency to $f$=200 Hz, we observe that the orientation of the caps and the black arrows begin to deviate from perpendicularity and particles are drawn together, regardless of their cap orientation (second panel in Figure \ref{fig:Condensation}e). Consequently, the particles form clusters in which they assume random orientations, as shown in the last panel of Figure \ref{fig:Condensation}e. The difference between $\theta_o$ and $\theta_v$ after switching to $f$=200 Hz is shown by the green data in Figure \ref{fig:Condensation}f. The distribution broadens to larger values, and the peak has shifted to around $\pi$/2 due to the downward sedimentation of the clustered Janus particles. Thus, the formation of motile clusters in this system appears unrelated to motion persistence or to the occurrence of collisions between the microswimmers, and instead appears to stem from attractive colloidal inter-particle interactions resulting from EHD flows around the Janus microswimmers, as discussed in more detail in Section \ref{Disc}.

\begin{figure}
    \centering
    \includegraphics[scale=.9]{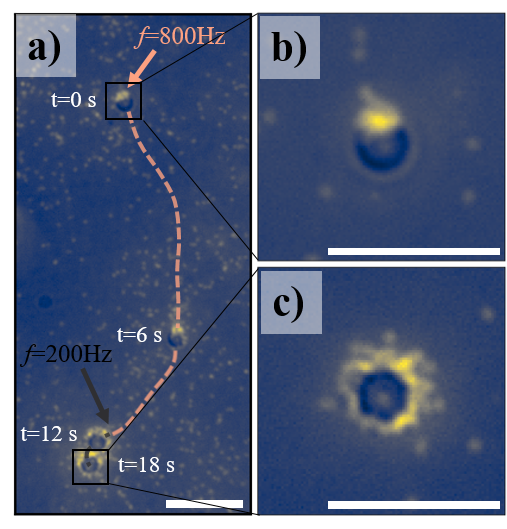}
    \caption{\textbf{Visualisation of electrically-induced flows through tracer particles. } a) Trajectory of a Janus microswimmer in the presence of fluorescent 2R = 200 nm PS tracers (yellow dots). The frequency is switched from $f$=800 Hz to $f$=200 Hz after 12 s. b \& c) Zoomed in snap shots of the microswimmer at t=0 s, $f$=800 Hz) (b) and t = 18 s, $f$=200 Hz (c). Scalebars: 10 $\mu$m}
    \label{fig:tracers}
\end{figure}

We probe the frequency-driven changes in the nature of the electrohydrodynamic (EHD) flows around the Janus microswimmers by adding small (2R=200 nm) polystyrene (PS) tracer particles to the experimental system (Fig. \ref{fig:tracers}). At $f$=800 Hz, the Janus microswimmer moves relatively quickly, pulling tracer particles (yellow) from the silica side and accumulating them at its rear (metallic cap)(Fig. \ref{fig:tracers}b). When the frequency is reduced to $f$=200 Hz, the microswimmer slows down, and the tracer particles become more isotropically distributed around the perimeter of the microswimmer (Fig. \ref{fig:tracers}c). This isotropic distribution suggests that the EHD flows become more uniform as the frequency decreases.

\balance
\section{Discussion} \label{Disc}

Our investigation reveals the complex dynamics and structural transformations of metallo-dielectric Janus microswimmers under confinement within a rectangular trap and subjected to a perpendicularly oriented AC electric field. By systematically adjusting the AC field frequency, we demonstrate that the self-propulsion of these microswimmers, as well as their assembly into varying structural forms, can be finely controlled. Specifically, we observe a frequency-dependent progression from a gas-like state of independent microswimmers to densely packed, motile crystallites exhibiting collective motion with pronounced vorticity as the frequency decreases from $f$=800 Hz to $f$=200 Hz.

In contrast to several previous studies employing metallo-dielectric Janus microswimmers in AC electric fields \cite{Yan2016, Nishiguchi2015, Demirors2018, Ganwal2008}, our experiments focus on the sub-kHz frequency range, where active motion persists at relatively low frequencies. This range aligns with the regime of EHD flows, driven by interactions between oscillatory electric fields and induced charges near the dielectric Janus particle's surface and the electrode. At lower frequencies $f \sim kD/H$, the motion is diffusion-dominated, while at higher frequencies $f \sim k^2D$, ions cannot respond quickly enough to the field. Within this intermediate range, i.e.  $\rm \kappa D/H \ll f \ll \kappa^2 D$, the electric field and induced charges generate forces that drive fluid motion \cite{Ristenpart2007}. 

In our system, this range corresponds to $\rm 439~Hz \ll f \ll 21720~Hz$. This implies that at frequencies below $\approx$500 Hz, the electric field period becomes significantly slower than the charge relaxation time in solution, leading to a strong reduction in the EHD flows. This explains the sharp decline in microswimmer self-propulsion velocity observed upon decreasing the frequency to $f$=200 Hz, as well as the tracer experiments (Fig.\ref{fig:tracers}a) showing a diminished contrast in the magnitudes of the EHD flow between the silica and metallic hemispheres of the microswimmers.

Clustering and crystallization under AC applied fields has been frequently reported in the literature for homogeneous dielectric microspheres across a broad range of frequencies \cite{Prieve2010}. At lower frequencies ($f$< 500 Hz) clustering may result from attractive lateral interactions driven by equilibrium-charge electroosmotic (ECEO) flows, also referred to as standard electroosmotic flows (EOF) \cite{Prieve2010, Bazant2004}. We indeed observe crystallization for bare silica spheres (i.e., without the Pd coating) across all explored frequencies (See Supplementary Figure S2) in contrast to the observations for our Janus particles reported above. Our results suggest that the relative importance of the different electrohydrodynamic flow components for Janus particles vary as a function of frequency. While at higher frequency self-propulsion generated by imbalances in EHD flows overcomes entrainment, at lower frequencies ECEO flows may prevail %due to the electric field acting on the equilibrium diffuse charge layer surrounding each microswimmer
, inducing flows that promote particle entrainment and aggregation irrespective of the orientation of the metallic cap.

\section{Conclusion}
In conclusion, our study demonstrates the tunable dynamics of dense monolayers of AC field-driven Janus microswimmers in the sub-kHz regime. We present a versatile experimental platform in which particle density is controlled, while microswimmer self-propulsion speed and inter-particle interactions can be tuned by varying the AC electric field frequency. At high frequencies, particles remain in a gas-like state, whereas decreasing the frequency drives their assembly into motile crystallites with emergent vorticity and collective motion. At low frequencies, the isotropic orientation of particles within these self-assembled crystallites indicates that long-range EHD flows dominate over alignment interactions, highlighting the potential of frequency-dependent EHD flows as a powerful tool for controlling clustering and internal structure in colloidal active matter systems.

By enabling dynamic control over both self-propulsion and assembly, our findings open up exciting possibilities for designing reconfigurable, high-density active systems with potential applications in smart materials, self-healing systems, and soft robotics. Additionally, our study lays the groundwork for further investigations into systems comprising mixtures of particles with varying self-propulsion speeds and inter-particle interactions. Finally, exploring the factors that govern the characteristic size of clusters at low frequencies will be a promising avenue for future research and the competition between the different components of electrohydrodynamic flows calls for further theoretical and numerical studies, as the low-frequency behavior of Janus active particles remains largely unexplored.

\section{Materials and Methods}
\subsubsection*{Janus particles}
Metallo-dielectric Janus microswimmers are fabricated by drop-casting an aqueous suspension of SiO$_2$ spheres (R=2.96 $\mu$m, SiO2-R-LSC84, microparticles GmbH) on a plasma-cleaned microscopy slide, followed by depositing a thin (2 nm) Cr adhesion layer and a 6.5 nm Pd layer. The resulting metallo-dielectric Janus particles were dispersed in 1 mL MilliQ water by ultrasonicating for 20 min., washed three times with a 1\% surfactant (Pluronic-127, Sigma Aldrich) solution by centrifugation and supernatant replacement, and finally concentrated by sedimentation and supernatant removal to yield a suspension of $\sim$2 mg particles/mL. Note that we do not remove the surfactant in order to prevent sticking to the electrodes of the experimental cell over the course of the experiment. Hence, experiments are carried out in the presence of 1\% Pluronic-127. 

\subsubsection*{Preparation of the experimental cell}\label{C4:SamplePrep}
Prior to each experiment, a particle traps was microfabricated on top of an indium tin oxide (ITO) coated glass substrate (resistivity$\approx 400 \Omega$) using two-photon polymerization of a photosensitive resin (Nanoscribe Photonic Professional GT, IP-S), followed by 20 minutes developing in propylene glycol methyl ether acetate (PGMEA, Sigma Aldrich), rinsing with isopropyl alcohol (IPA, Sigma Aldrich), and finally one min. exposure to UV light (365 nm @ 3 W/cm$^2$, Nitecore GEM10UV) to ensure full curing of the structure. To assemble the experimental cell, a round, $h$=30 $\mu$m thick double sided adhesive spacer (No.5603, Nitto) was placed around the microfabricated structure. The resulting well was filled with 1.5 uL of the Janus particle suspension, and finally, the cell was closed by sticking another ITO coated substrate on top of the assembly (ITO coated side down). The ITO-coated substrates, i.e. the electrodes, were then connected to a function generator (National Instruments Agilent 3352X, USA) that generates an AC electric field with a frequency ranging from $f$ = 200-800 Hz, and an applied peak-to-peak voltage of $V_{pp}$ = 4 V, corresponding to a field of $E=V_{pp}/h\approx$133 V/mm, unless stated otherwise. The entire sample cell was placed at an angle of $\alpha$=45$^\circ$ degrees with respect to the optical table, and imaged in bright-field transmission mode using a custom-build optical setup comprising an infinity-corrected low magnification (M PLAN APO 20x, Optem) or high magnification (Plan Apo Infinity Corrected 50x, Mitutoyo) long-working-distance objective connected to a CMOS camera (Orca Flash4.0 V3, Hamamatsu). Images of 2048x2048 pixels were taken at a frame rate of 30 frames per second. 

\subsection{Data Analysis.}\label{C4:MM_DataAnalysis}
From the experimental videos, particles were tracked using a centroid finding algorithm implemented in Python (Trackpy \cite{Crocker1996}). 

\subsubsection*{Hexagonal order parameter}
From the obtained particle coordinates, we determine the hexagonal order parameter, $\psi_6$, and subsequently the phase of $\psi_6$ as follows. First, we identify for each particle $i$ the number of nearest neighbors in the first shell ($N$), and calculate the angles between the reference particle and each of the $j$-th neighbors, $\theta_j$. The value of $\psi_6$ is then calculated by taking the average of the complex exponential of six times these angles: 
\begin{equation}
    \psi_6 = \frac{1}{N} \sum^N_{j=1}e^{i\cdot6\theta_j}
\end{equation}
which yields a complex number indicating the local hexagonal order. In particular, the absolute value of $\psi_6$ (denoted as |$\psi_6$|) indicates the degree of hexagonal ordering around the particle. Additionally, the phase of $\psi_6$ provides information about the local orientation of the hexagonal ordering around each particle and is calculated as: 
\begin{equation}
    phase(\psi_6) = arg(\psi_6)
\end{equation}
Here, $arg()$ denotes the argument (or angle) of the complex number in the complex plane.

\subsubsection*{Vorticity}
The vorticity of the microswimmers was computed by first calculating the velocity gradients and then deriving the vorticity from these gradients. First, we identify for each particle $i$ the neighbors $j$ in the first two surrounding shells. Then, assuming the displacement of the neighbors with respect to particle $i$ is given by the sum of a translation and a rotation, the velocity gradients were derived by solving the system: 
\begin{equation} \label{eq:matrixSystem}
    B = A \cdot \textbf{J}
\end{equation}
Where $B = \begin{psmallmatrix} x_j & y_j & 1\\ \cdot & \cdot & 1 \\ \cdot & \cdot& 1 \\ \cdot & \cdot & 1 \end{psmallmatrix}$
are the initial positions of the neighbors augmented with an additional column of ones to facilitate the inclusion of translational components. The matrix $A$ = $\begin{psmallmatrix} \Delta x_j \\ \Delta y_j \\ \cdot \\ \cdot \\ \cdot \end{psmallmatrix}$ contains the displacements after time step $\Delta t$, and $\textbf{J}$ = $\begin{psmallmatrix} \frac{\delta v_x}{\delta x} & \frac{\delta v_x}{\delta y} \\ \frac{\delta v_y}{\delta x} & \frac{\delta v_y}{\delta y}\end{psmallmatrix}$ is the Jacobian matrix. To obtain the velocity gradients, eq. \ref{eq:matrixSystem} was solved for $\textbf{J}$ by calculating the dotproduct of $B$ and the inverse of matrix $A$, $A^+$. Finally, the vorticity $\omega$ was calculated as the difference between the partial derivatives of the velocity components: 
\begin{equation}
    \omega = \frac{\delta v_y}{\delta x} - \frac{\delta v_x}{\delta y}
\end{equation}

\subsubsection*{Janus microswimmer cap orientation}
The cap orientation of individual microswimmers was extracted from high-magnification images using a custom code written in Matlab. In brief, we first find the center of each particle using a centroid finding algorithm \cite{Crocker1996}. We then apply a bandpass filter to the original image with a passband frequency range of 2-7 pixels, which yields the caps of the Janus particles to appear as bright blobs. Subsequently, circular regions were cropped around the particle center, and the brightness center was obtained by weighing the intensity of the pixels in the corresponding crop with the distance from the center position. Finally, the center and the brightness center were connected to obtain the orientation of the two-dimensional Janus particle.

\section*{Author Contributions}
Author contributions are defined based on the CRediT (Contributor Roles Taxonomy) and listed alphabetically. Conceptualization: L.A., C.v.B. Formal Analysis: C.v.B. Funding acquisition: L.I. Investigation: L.A., C.v.B. Methodology: L.A., C.v.B. Project Administration: L.I., Software: C.v.B. Supervision: L.I. Validation: C.v.B. Visualization: C.v.B Writing - original draft: C.v.B. Writing - review and editing: L.A., C.v.B. .

\section*{Conflicts of interest}
The authors declare that they have no conflicts of interest.

\section*{Acknowledgements}
We acknowledge Roberto Piazza, Stefano Buzzaccaro, Chantal Valeriani and Nuno Araujo for useful discussions, and Moritz Röthlisberger for preliminary experiments. C.v.B. acknowledges funding from the European Union’s Horizon 2020 MSCA-ITN-ETN, project number 812780. L.I. acknowledges funding from the European Research Council (ERC) under the European Union’s Horizon 2020 Research and Innovation Program grant agreement No 101001514.

%%%END OF MAIN TEXT%%%


\balance

\renewcommand\refname{References}

%%%REFERENCES%%%
\bibliography{bib} %You need to replace "rsc" on this line with the name of your .bib file
\bibliographystyle{rsc} %the RSC's .bst file

\end{document}

\baselineskip24pt

\maketitle 
\newpage
\subsection*{S1. Vertical density profiles}
\begin{figure}[h!]
\renewcommand{\thefigure}{S1}
    \centering
    \includegraphics[width=\textwidth]{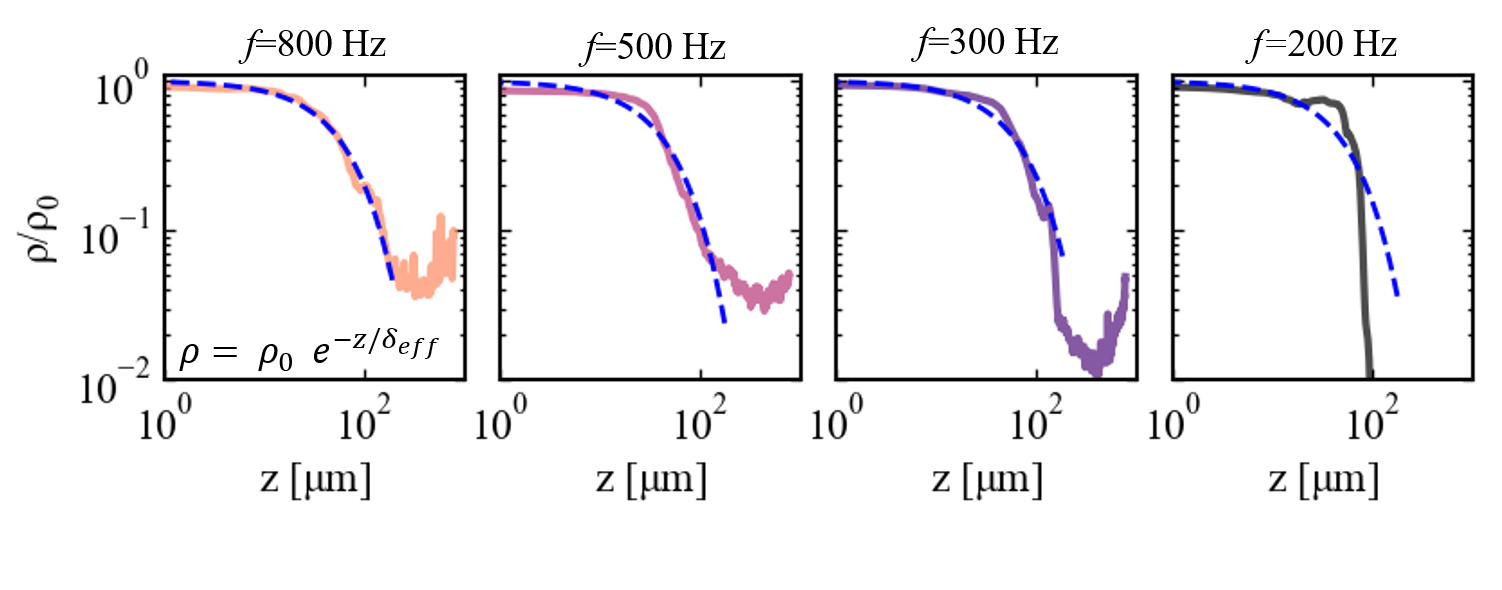}
    \caption{\textbf{Vertical density profile of Janus microswimmers at different frequencies.} Average number density profile as a function of height of Janus microswimmers at different frequencies and $V_{pp}$ = 4 V, taken from the 400 $\mu$m center of the particle trap after 9 min. of equilibration. Blue dotted lines indicates best fits to ideal gas law, $\rho = \rho_0 e^{-z/\delta_{eff}}$. 
    }
    \label{fig:VerticalDensity}
\end{figure}

At $f$=800 Hz, our system shows features analogous to the active gas-like state observed for suspension of Pt-capped catalytically active Janus particles under a gravity field \cite{Palacci2010,Ginot2015, Ginot2018}. Figure \ref{fig:VerticalDensity} shows normalized average density profiles as a function of height for different frequencies of the AC electric field. Blue dotted lines show fits to the ideal gas law, $\rho = \rho_0 e^{-z/\delta_{eff}}$, where $\delta_{eff} = \frac{k_B T_{eff}}{m^* g sin\theta}$ corresponds to the effective sedimentation length, with $T_{eff}$ the effective temperature, $m^*$ the buoyant mass of a single microswimmer, and $g sin\alpha$ the effective gravitational constant at the angle $\alpha=45^\circ$. At $f$=800 Hz the data follow the shape of the ideal gas law in the lower part of the trap. However, the effective temperature we extract from the sedimentation length is slightly lower than the effective temperature extracted from the effective diffusion coefficient of the microswimmers ($\sim 10^5$ K versus $\sim 10^6$ K, respectively), suggesting the importance of other factors such as inter-particle interactions. In contrast, at higher $f$ the fit is not an accurate description of the data.

\newpage

\subsection*{S2. Comparison of Pd capped and bare SiO$_2$ spheres}

\begin{figure}[h!]
\renewcommand{\thefigure}{S2}
    \centering
    \includegraphics[width=\textwidth]{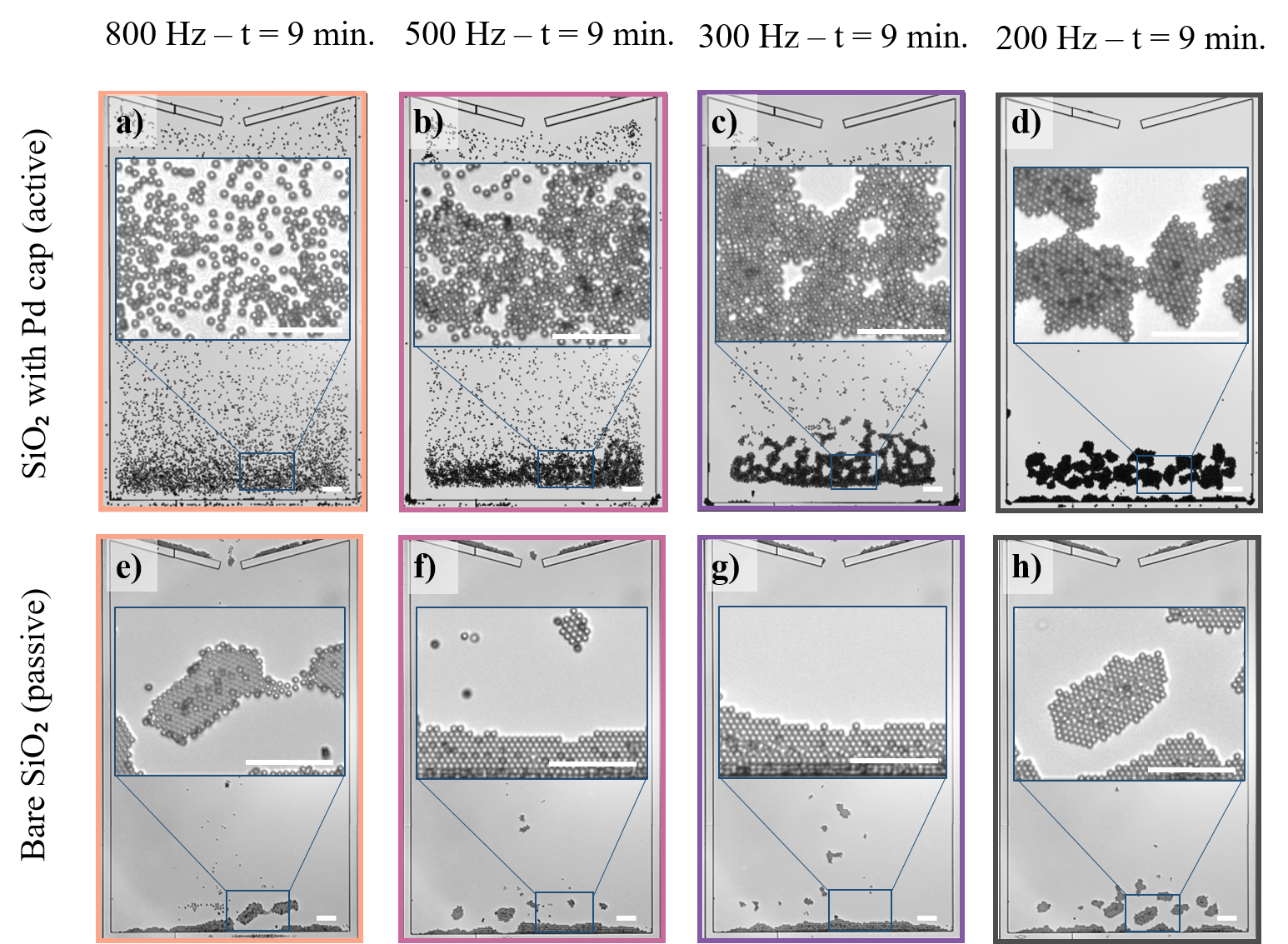}
    \caption{\textbf{Comparison of Pd capped and bare SiO$_2$ spheres.} Snapshots of the system described in the main text after 9 min. of equilibration at different AC electric field frequencies at a peak-to-peak voltage of $V_{pp}$ = 4 V for (a-d) active Janus particles (SiO$_2$ spheres with a Pd cap) and (e-h) passive particles (uncoated SiO$_2$ spheres). Scale bars: 50 $\mu$m. 
    }
    \label{fig:Passive_vs_Active}
\end{figure}

\clearpage

\subsection*{S3. Time scale estimation}

The relaxation time scales of ions in solution under the effect of an AC electric field have been extensively studied by Squires, Bazant, Ristenpart and Delgado in previous works \cite{Bazant2004, Squires2004, Ristenpart2007}. To calculate the range of frequencies at which electrohydrodynamics (EHD) flows take place we have taken into account different parameters. In our solution, the conductivity and dielectric constant of the media are $\rm \sigma_m =1.5~\times~10^{-5}~Sm^{-1}$ and $\rm \epsilon_m= 78$ (25$^{\circ}C$), respectively. Thus, the Debye length $\rm \kappa^{-1}$ is estimated as 

\begin{equation}
    \rm \kappa^{-1}= \sqrt{ \frac{\epsilon_m \epsilon_0 D}{\sigma_m}},
\end{equation}\\

with $\rm \epsilon_0 = 8.854~\times~10^{-12} Fm^{-1}$, $\rm D = 2~\times 10^{-9} m^{2}s^{-1}$ considering a general value for ions in the media to be able to do the calculation. Thus, in our system, $\rm \kappa^{-1} \approx 300 nm$. A summary of all the parameters used in the calculations in the main text and supplementary information are depicted in Table.\ref{tab:param}

\begin{center}
\begin{table}[h]
\renewcommand{\thetable}{S1}
    \centering
    \begin{tabular}{c|c}
    \hline
      Parameters   &  Values \\
    \hline
        H & $\rm 1.5\times10^{-5}~[m] $\\    
        $\rm V_{pp}$ & 4  \\    
        $\rm \sigma_m$ & $\rm 1.5\times10^{-5}~[Sm^{-1}]$ \\   
        $\rm \epsilon_m$ & 78 \\
        $\rm \epsilon_0$ & $\rm 8.854\times10^{-12} [Fm^{-1}]$ \\ 
       D & $\rm 2\times 10^{-9} [m^{2}s^{-1}]$  \\    
        \hline
    \end{tabular}
    \caption{ Parameters used in the calculation of the various timescales and other quantities in this system.}
    \label{tab:param}
\end{table}
\end{center}

\section*{List of Supplementary Videos}
\begin{itemize}

\item \textbf{Video S1}: Video of the metallo-dielectric Janus particles confined in a rectangular trap under gravity after 9 minutes of equilibration under and AC electric field with frequency of 800 Hz and  a peak-to-peak voltage of $V_{pp}$ = 4 V. Video corresponds to the snapshot in Figure 1e in the main text. The video is sped up 2 times. 

\item \textbf{Video S2}: Video of the metallo-dielectric Janus particles confined in a rectangular trap under gravity after 9 minutes of equilibration under and AC electric field with frequency of 500 Hz and  a peak-to-peak voltage of $V_{pp}$ = 4 V. Video corresponds to the snapshot in Figure 1f in the main text. The video is sped up 2 times. 

\item \textbf{Video S3}: Video of the metallo-dielectric Janus particles confined in a rectangular trap  under gravity after 9 minutes of equilibration under and AC electric field with frequency of 300 Hz and  a peak-to-peak voltage of $V_{pp}$ = 4 V. Video corresponds to the snapshot in Figure 1g in the main text. The video is sped up 2 times.

\item \textbf{Video S4}: Video of the metallo-dielectric Janus particles confined in a rectangular trap  under gravity after 9 minutes of equilibration under and AC electric field with frequency of 200 Hz and  a peak-to-peak voltage of $V_{pp}$ = 4 V. Video corresponds to the snapshot in Figure 1h in the main text. The video is sped up 2 times. 

\item \textbf{Video S5}: Condensation of microswimmers under gravity by switching the frequency of the AC electric field from 800 Hz to 200 Hz at a fixed peak-to-peak voltage of $V_{pp}$ = 4 V. Microswimmers are color-coded according to their cap orientation. Black arrow indicated their direction of motion extracted from their velocity vector. Video corresponds to the snapshots in Figure 4e in the main text. The video is real time. 

\end{itemize}

\clearpage

\bibliography{bib} %You need to replace "rsc" on this line with the name of your .bib file
\bibliographystyle{rsc} %the RSC's .bst file